# Nitric Oxide in Climatological Global Energy Budget During 1982–2013

Cissi Y. Lin[1]

Yue Deng[1]

[1]Department of Physics, University of Texas at Arlington, Arlington, Texas, USA

Corresponding author: yuedeng@uta.edu

**Abstract**

Over the past decades, temperature and density of the upper atmosphere show negative trends and decrease of the upper atmospheric temperature is attributed to the declining neutral density. Specifically, nitric oxide (NO) and carbon dioxide ($CO_2$) govern thermospheric cooling at 5.3 µm and 15 µm, respectively. While a lot of efforts have focused on the $CO_2$ effects on the long-term trends, relatively less attention has been paid to the impacts by NO, which responds to solar and geomagnetic activities dynamically. In this study, we investigate the role of NO in climatological global energy budget for the recent three solar cycles using the Global Ionosphere Thermosphere Model. From 1982 to 2013, the F10.7 and Ap indices showed a decadal decrease of ~8% and ~20%, respectively. By imposing temporal-varying F10.7 and Ap values in the simulations, we find a decadal change of -0.28x$10^{11}$ W or -17.3% in total NO cooling power, which agrees well with that (-0.34x$10^{11}$ W or -17.2%) from the empirical Thermosphere Climate Index derived from the TIMED/SABER data. Neutral density decreases by 10–20% at 200–450km and $T_{ex}$ decreases by 25.3 K per decade. The deduced-decadal change of NO cooling reaches ~25% of the sum of total heating at ~130 km and its significance decreases with altitude.

**Plain Language Summary**

Temperature and neutral density of the upper atmosphere has declined over the past decades. While the increase of $CO_2$ in the lower atmosphere resulting from mankind activities is attributed to cause the decreases through cooling the upper atmosphere, the declining solar and geomagnetic activity also play a role. However, the abundance of the other cooling agent, NO, reflects the variability in solar and geomagnetic activity and therefore compete against the trend brought by $CO_2$. Using a global 3D model, we investigate the role NO plays in the long-term energy budget for the past 32 years.

Keywords: nitric oxide, thermospheric energy budget, cooling

**Key Points**

1. For 1982–2013, GITM total NO cooling and TCI display a decadal decrease of -17%.

2. The variations of modeled density and temperature agree with other measurements.

3. At ~130 km, the magnitude of NO cooling decrease is ~25% of that of total heating decrease.

# 1 Introduction

Over the past decades, temperature and density of the upper atmosphere show negative trends [Keating et al., 2000]. Empirically removing the solar and geomagnetic factors, Emmert [2015] inferred a decadal trend of -1% to -5% from the derived anomaly in the thermospheric neutral density and that of up to -6 K in exospheric temperature. Table 1 of Emmert [2015] and Figure 4 of Solomon et al. [2015] summarize decadal variations ranging from -1% to -7% from these recent studies. One of the major cooling mechanisms in the upper atmosphere is via the 15-µm infrared emission by carbon dioxide ($CO_2$). As $CO_2$ level has consistently risen over the past decades owing to the terrestrial mankind activities, this greenhouse gas species has been the focus of decades of extensive studies [Roble and Dickinson, 1989; Qian et al., 2006, 2011, 2015; Solomon et al., 2015]. The impact of solar and geomagnetic variability to the upper atmospheric was considered in these modeling efforts so that the derived long-term change may be conclusively explained by the increasing amount of $CO_2$ from the lower atmosphere. A similar removal attempt, but with a different technique, of the solar and geomagnetic components has been applied to the neutral temperature derived from incoherent scatter radars (ISR) measurements. A decadal trend of -10–15 K [Ogawa et al., 2014] and -18 K (-69 K over 38 years in [Oliver et al, 2014]) were observed at Tromsø and Millstone Hill, respectively. The discrepancies between the trends derived from the ISR measurements and those from the model and derived from satellite drag derived temperature were concluded as results of different increasing rate of $CO_2$ [Ogawa et al., 2014] with altitudes and variations in heating and cooling in the thermosphere [Oliver et al, 2014].

Relatively less attention has been paid to nitric oxide (NO), the second major thermospheric cooling channel at 5.3 µm, for its climatological behavior owing to the complexity of the role NO plays in the thermosphere. NO is a key species in the lower thermospheric photochemistry though being a minor constituent. Its production is sensitive to those energy sources able to break the strong bond of molecular nitrogen and its abundance is indicative of solar and geomagnetic energy deposition [Baker et al., 2001; Barth and Bailey, 2004; Barth et al., 1988; 1999; 2001; 2003; Siskind et al., 1990]. When colliding with O, NO is excited to the vibrational level (v = 1). The vibrational (1–0) transition removes the kinetic energy and emits in the 5.3-µm band. The amount of emission, radiating to cool the atmosphere, is a function of NO and O density as well as the ambient temperature. Both NO abundance and emission display geophysical signatures of various spatial and temporal scales, such as solar cycles [Kumar et al., 1999], solar rotation [Barth et al., 1999], flares [Rodgers et al., 2010] and geomagnetic storms [Baker et al., 2001]. Today, it remains a challenge to accurately reproduce both NO density and cooling in models.

An updated photochemistry scheme has been performed recently to the Global Ionosphere-Thermosphere Model (GITM) to provide a suitable photochemistry for climatological NO studies. In that work, quantitative changes of NO density and cooling owing to the chemistry update and the sensitivity of global NO cooling to solar and geomagnetic forcing are investigated. To improve our understanding of the energy budget in the upper atmosphere, especially over the past few decades, it is critical to address the questions such as: 1) What is the long-term variation of NO cooling? and 2) Climatologically, what is the significance of NO cooling compared to the long-term variabilities of other mechanisms? In this study, we perform 3D global simulations with the updated photochemistry scheme for 1982–2013 and examine the total NO cooling power (essentially global integrated emission at 5.3 µm). The consequential changes in neutral density and thermospheric temperature are extracted at three satellite altitudes, – Challenging Minisatellite Payload (CHAMP), Gravity Recovery and Climate Experiment (GRACE), and Gravity Field and Steady-State Ocean Circulation Explorer (GOCE) –, to make available for model-observation comparisons. The significance of NO cooling in the long-term thermospheric energy budget is further examined. To our knowledge, the long-term variation of NO cooling and its potential impacts to the upper atmosphere have not been closely examined previously. It is, therefore, critical to reproduce it well and recognize the role it plays in the long-term climate with a self-consistent upper atmospheric model.

**2 Methodology**

Based on the 15-year Sounding of the Atmosphere using Broadband Emission Radiometry (SABER) measurements, Thermosphere Climate Index (TCI), parameterized through the F10.7, Ap, and Dst indices, was proposed as an empirical proxy for global NO emission [Mlynczak et al., 2016]. The coefficients of the empirical equation were later updated along with the provision of the parameterization for the global SABER $CO_2$ emission through the same indices. The empirical formulas provide an efficient way to estimate global infrared emission by both species since 1947 and TCI can be further used to describe the thermal level of the global thermosphere Mlynczak et al. [2018]. Interestingly but not surprisingly, during the past three decades when the solar activities are known to decline, the empirically calculated emission of both species show a downward tendency. Using the empirical equations [Mlynczak et al., 2016], the global emission by $CO_2$ and NO for 1982–2013 shown in Figure 1(a) suggests decadal decreases of 5% and 17%, respectively. The period of 1982–2013 is selected to cover nearly equal numbers of years of high and low solar activities.

Figures 1(b) and 1(c) show the daily and yearly averages of F10.7 and Ap indices, respectively. A linear fit to each data set shows a similar downward slope regardless using daily or yearly averages as the linear fit only captures the large-scale temporal variation. Over the presented period, the F10.7 and Ap indices have decreased by -8.1% and -19.0%, respectively. Decreases of solar energy input generally result in lower exospheric temperature, $T_{ex}$, and lower neutral and electron densities. For 1982–2013, we derived a decadal decrease of ~15% in neutral density using the satellite drag data (provided by Emmert [2015]). During the anomalously low solar minimum in 2008, the F10.7 index and neutral density is 3.7% and 28% lower than the previous solar minimum respectively [Emmert el al., 2010]. Note that these values regard the data prior removing the solar and geomagnetic factors. After the empirical fitting has been removed from the data, the residual shows that the anomaly posed by the elongated solar minimum during 2007–2009 is strong enough to offset the derived long-term decadal trends by 0.4 K in $T_{ex}$ and 0.8% in neutral density for 1967–2013 compared to the trends derived from 1967–2005 [Emmert, 2015]. On the other hand, the terrestrial atmosphere appears to intensify its response (in terms of the magnitude of decline) during anomalously low and prolongated solar minima of cycle 23/24 even more than model runs with the varying solar extreme ultra-violet irradiance, geomagnetic forcing, and $CO_2$ cooling considered [Solomon et al., 2011].

NO participates actively in the chains of photochemical reactions and energy budget in the lower thermosphere. Its density is a direct reflection of the energy deposition to the lower thermosphere. While its density increase tends to be followed by its own enhanced cooling, the consequential ambient temperature determines its rates of production and loss. Owing to the fact that NO density and cooling are sensitive to the solar and geomagnetic conditions, the long-term decrease of F10.7 and Ap indices in the last several decades likely leads to a long-term variation in both NO density and cooling. Since solar and geomagnetic forcing is the primary driver for the NO variation, separating NO calculation from being impacted by solar and geomagnetic forcing will violate the self-consistency of a model. Therefore, the climatology of solar and geomagnetic conditions should not be excluded when studying the climatology of NO cooling.

GITM has been driven with the annual averages of F10.7 and Ap for the 32 years during 1982–2013. Each year is simulated separately for a period of five days and NO cooling is accounted when the simulation reaches a quasi-steady state after imposing the corresponding solar and geomagnetic conditions. GITM only considers NO cooling as the removal of kinetic energy from the atmosphere through the vibrational (1–0) transition [Kockarts, 1980] with the quenching rate with O of $2.59 \times 10^{-11}$ cm3 s-1 [Caridade et al., 2008]. NO volume cooling rate is integrated over the globe to obtain total cooling power.

## 3 Results

Figure 2(a) shows the modeled total NO cooling power for the period of 1982–2013 in red. The TCI values shown in blue are calculated with the corresponding annual-average indices. The dashed lines show linear fits of the corresponding data sets. The linear fit of GITM cooling power shows a decadal decrease of $-0.29 \times 10^{11}$ W or -17.1%. The linear fit of TCI shows a decadal decrease of $-0.34 \times 10^{11}$ W or -17.2%. Both of them are highly correlated with F10.7: 99.2% for GITM and 97.7% for TCI. The correlation with Ap is slightly lower: 66.4% for GITM and 75.5% TCI. This does not come as a surprise. We have shown and discussed in our sensitivity study that GITM total cooling power responses more strongly to changes in F10.7 than those in Ap as averagely solar radiation has a greater impact to global NO. Figure 2(a) shows that in general total cooling power from GITM agrees with TCI better during the ascending phase of a solar cycle than during the descending phase and better at higher solar activity than at lower solar activity. Particularly, the model agrees well with TCI during the peak years of Solar Cycle 22 (1989–1992) and the quiet year of 2009. Figure 2(b) shows the year-to-year variation, defined as the percentage of variation froma year prior to the current plotted year. In general, the variationss of the modeled cooling power and TCI agree well. GITM has a steeper year-to-year variation. As shown in Figure 2(a), the model overall has lower total cooling power during solar minimum and rises to a similar level of TCI during solar maximum. This tendency results in the more positive year-to-year variation ascending to solar maximum and less negative variation descending to solar minimum.

The declining activity from the Sun (in F10.7 and Ap) can be observed in the modeled neutral density and exospheric temperature, $T_{ex}$. Figures 2(c–e) show the neutral density extracted at 450, 300, and 250 km, corresponding to the average altitudes that CHAMP, GRACE, and GOCE spacecraft make measurements. All of them display a decadal change during 1982–2013 of: -19.8% at 450 km, -16.9% at 300 km, and -13.0% at 250 km. These values agree well with what was observed in the neutral density derived from the satellite drag data during the same period of time before background removal. Using the neutral density derived from the satellite drag data during the same period of time before background removal (supporting data in Emmert [2015]), we obtain decadal changes of -19.9% at 475 km, -14.8% at 325 km, and -11.4% 250 km for 1982–2013. Figure 2(f) shows that the $T_{ex}$ has a decadal change of -25.3 K. The decreases mainly result from the collapse of the upper atmosphere owing to the decrease of the solar activity. Solomon et al. [2011] reported up to -60 K of difference in the modeled exospheric temperature of the recent two solar minima. We deem our findings within a reasonable range of agreement.

The two thermospheric cooling agents have a strong dependence on the solar cycle phase: $CO_2$ dominates the variation at solar minima and NO dominates that at solar maxima. To provide more perspective, we also derive long-term variations for both solar minima and maxima. At minima, the decadal change is -15.9% and -21.8% for GITM cooling power and TCI, respectively.

A glance at resulting neutral density at 400 km and $T_{ex}$ at solar minima shows decadal changes of -19.7% and -5.7 K. Our findings agree well with Emmert [2015], in which the implicit density trends by the F10.7 and Kp indices were reported to contribute to a neutral density change of -13.9%. At maxima, the decadal change is -27.5% for GITM cooling power and -25.3% for TCI, resulting changes of -32.1% and -77.5 K in neutral density at 400 km and $T_{ex}$. It has been shown that GITM cooling power is less/more sensitive to Ap than TCI at lower/higher solar activity. Since the annual average of F10.7 is around 70 (74, 72, and 71) for the three minima, the variation of cooling power is dominated by that of Ap. The average F10.7 for the three maxima are 190, 181, and 123. As a result, the variation of GITM cooling power mainly display its sensitivity to the level of solar forcing and close to the TCI variation.

For the purpose of further investigation into the thermospheric energy budget, the years 1988 and 2011 are selected out of the six years where the linear fit intersects with the total cooling power in Figure 2 because they are both in the ascending phase of a solar cycle and are farthest apart to be representative for the long-term variations we observe in Figures 2 and 3. The differences between the two years are divided by 2.3 to provide an estimate for decadal variabilities. Hereafter, we will explicitly use *deduced-decadal* to describe the quantities that are obtained using this approach. Figure 3(a) shows the vertical profile of the deduced-decadal change in global-mean thermospheric temperature. A negative tendency is visible throughout all altitudes and reaches -30 K at the isothermal altitudes. The magnitude is slightly greater than that from the linear fit (-25.3 K) when the temperature tendency profile is derived from these two years. It is consistent with the results shown in Figure 2(f) as $T_{ex}$ of 1988 is slightly above where the linear fit and $T_{ex}$ of 2011 is slightly below the linear fit. On the other hand, as the lower-boundary conditions are driven by the Naval Research Laboratory Mass Spectrometer Incoherent Scatter Radar (NRLMSISE-00) model [Picone et al., 2002] and the trend brought by $CO_2$ from the lower atmosphere is not included, the combination of slightly higher F10.7 and slightly lower Ap results in a non-zero but subtle temperature difference at 100 km. Figure 3(b) shows the vertical profiles of global-mean total heating (positive terms in the energy equation) for 1988 and 2011. Figure 3(c) shows the negative deduced-decadal change of total heating in red. NO cooling also has a negative tendency as shown in Figure 2(a) and a *decrease of NO cooling* is equivalent to a *positive change in heating* shown in blue. Figure 3(d) shows the ratio of the magnitude of NO cooling change (blue line in 3(c)) to that of total heating change (red line in 3(c)). The role that NO cooling plays, particularly in decadal change of energy budget, is most significant at 100–200 km, where NO is most abundant. At ~130 km, the magnitude of the deduced-decadal NO cooling variation is about 25% of that of total heating, which indicates the long-term NO cooling variation is not negligible in the long-term variation of the total energy budget and its effect may feedback to the long-term temperature change and reduce the temperature decrease in the long-term variation. Meanwhile, the long-term NO cooling decrease may result in steeper temperature growth with altitudes if other factors kept the same at those altitudes. Above 220 km, the NO cooling tendency is less than 5% of that of total heating. In summary, the declining geophysical factors (namely, F10.7 and Ap) from 1988 to 2011 result in less heating as well as less NO cooling (owing to lower NO density). The combined impact of decreases in both heating and NO cooling from the declining solar activity results in the overall $T_{ex}$ decrease.

The effect of increasing lower thermosphric $CO_2$ abundance has not been considered in this study. Higher $CO_2$ concentration contributes more cooling and drives the tendency toward temperature decrease – consequently lowering NO density and cooling. The consequential increase by NO 'heating' (from decrease of NO cooling) competes with the increase of $CO_2$ cooling. A future paper will include $CO_2$ variation in the scenario and discuss the roles $CO_2$ and NO play in the long-term thermospheric energy budget.

## 4 Summary

Global simulations have been performed for 1982–2013 with temporal-varying solar and geomagnetic indices, which are the main drivers to NO variation. Total NO cooling power shows strong correlation with solar activity level. The declining geophysical forcing results in decreases in total heating as well as NO cooling. The model results show a decadal change of -0.28x10$^{11}$ W or -17.3% in total NO cooling power, agreeing well with TCI (-0.34x10$^{11}$ W or -17.2%). Neutral density decreases by 10–20% at 200–450km and $T_{ex}$ decreases by 25.3 K per decade. The decadal changes derived from three minima are -15.9% for neutral density and -5.7 K for $T_{ex}$. The decadal changes derived from three maxima are -27.5% for neutral density and -77.5 K for $T_{ex}$. The decadal change of NO cooling reaches ~25% of the sum of all other heating terms below 150 km and its significance decreases with altitude. The decrease of overall heating causes $T_{ex}$ to decrease.

**Acknowledgement**

This research at the University of Texas at Arlington was supported by NSF through grant ATM0955629, NASA through grants NNX13AD64G and NNX14AD46G, and AFOSR through award FA9550-16-1-0059 and MURI FA9559-16-1-0364. The authors acknowledge the Texas Advanced Computing Center (TACC) at the University of Texas at Austin for providing Lonestar5 and Maverick resources that have contributed to the research results reported within this paper. URL: http://www.tacc.utexas.edu. The GITM model outputs are available at https://github.com/cissilin/Lin2018_NO_Long-Term.

Solomon, S. C., L. Qian, L. V. Didkovsky, R. A. Viereck, and T. N. Woods (2011), Causes of low thermospheric density during the 2007–2009 solar minimum, J. Geophys. Res., 116, A00H07, doi:10.1029/2011JA016508

Solomon, S. C., L. Qian, and R. G. Roble (2015), New 3-D simulations of climate changeinthethermosphere,J.Geophys. Res. Space Physics, 120, 2183–2193, doi:10.1002/2014JA020886

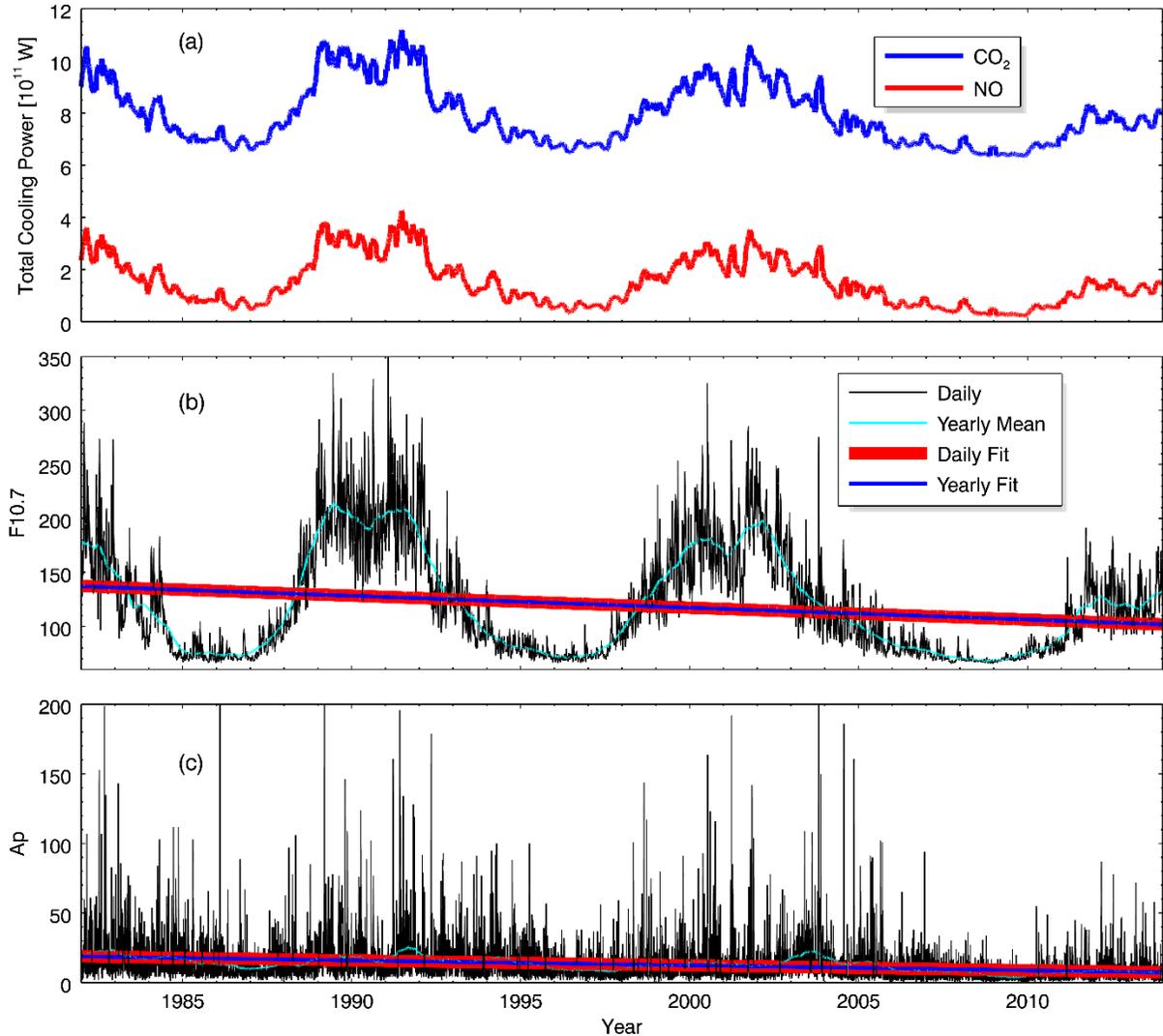

Figure 1. (a) Global infrared emission estimates using the Mlynczak et al. [2016] parameterization, (b) F10.7, and (c) Ap indices from Jan 1, 1982 to Dec 31, 2013. The linear fit of the daily (black) and yearly indices (light blue dotted) is shown as thick red and thin blue lines.

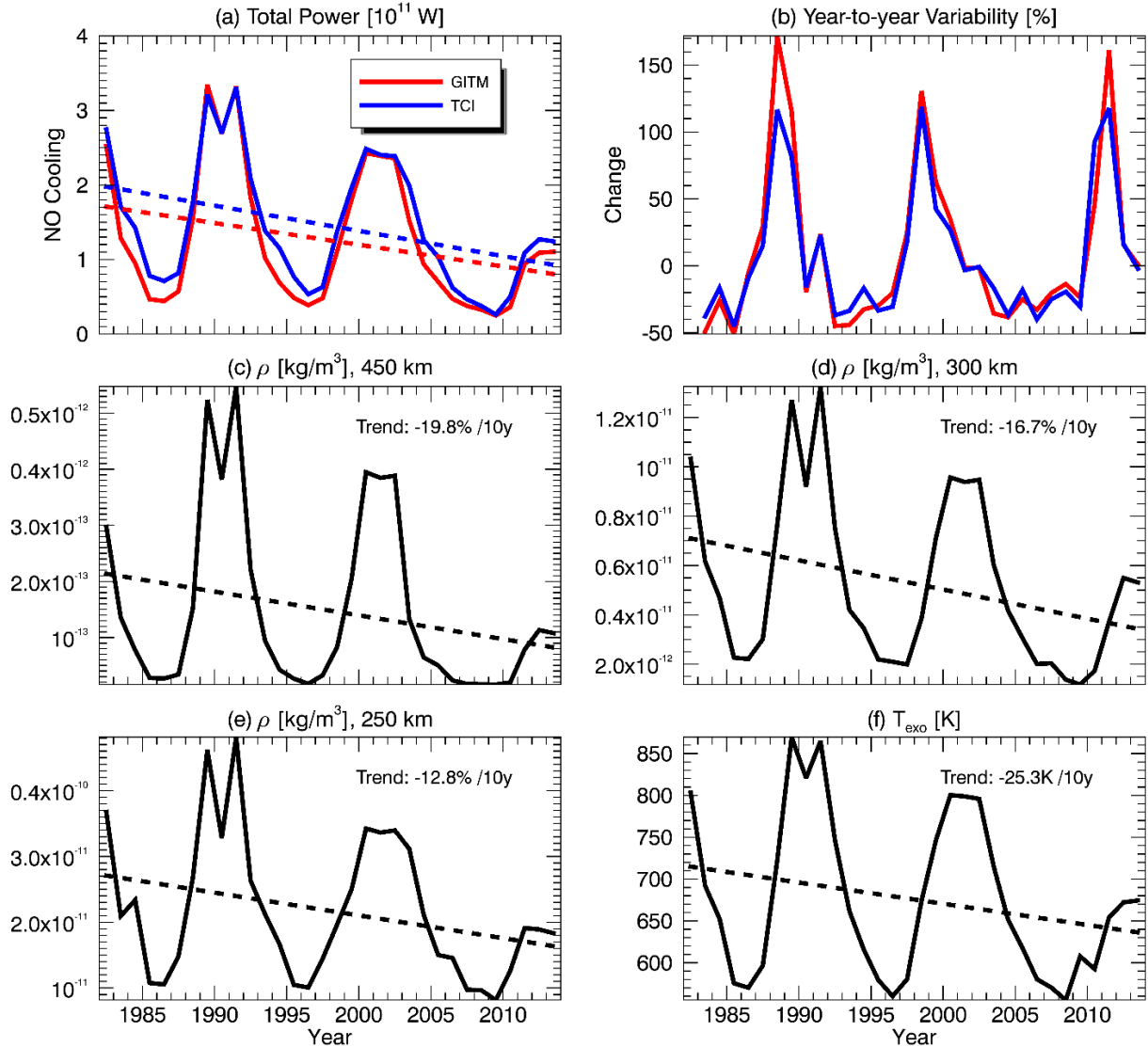

Figure 2. Model results with yearly average of F10.7 and Ap for 1982–2014: (a) Total GITM NO cooling power (solid red) compared with TCI (solid blue). The dashed lines are linear fits of the corresponding data sets. (b) Yearly variation is defined as the percentage of variation from the value a year prior to the current plotted year. The thermospheric variabilities are observable in neutral mass density at (c) 450, (d) 300, and (e) 250 km and in (f) exospheric temperature. The dashed line shows the linear fit to each case.

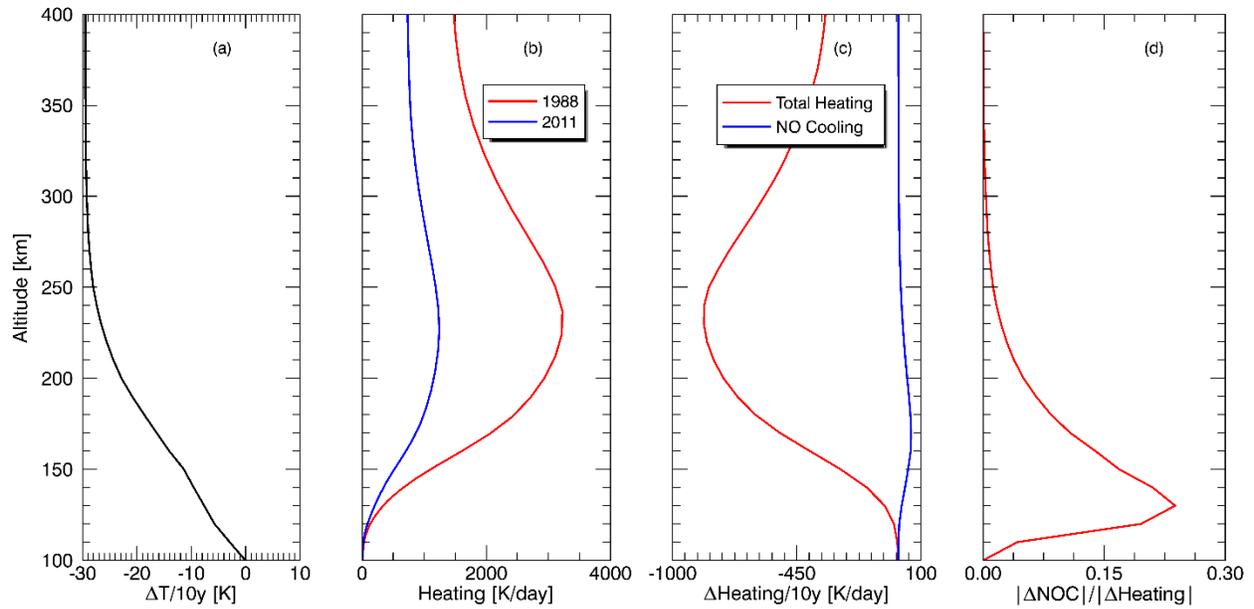

Figure 3. (a) Decadal change in thermospheric temperature. (b) Total heating terms for 1988 and 2011. (c) Deduced-decadal trend of total heating in red and that of NO cooling, a positive change in heating, in blue. (d) Ratio of the magnitudes of NO cooling trend to total heating trend.